\documentclass{article}
\usepackage{graphicx}  
\usepackage{amsmath}   
\usepackage[compress]{cite}
\usepackage{amssymb}   
\usepackage{bm} 
\usepackage{dcolumn}
\usepackage{color}
\usepackage{mathrsfs}
\usepackage{amsfonts}
\usepackage{varioref}
\usepackage{float}
\restylefloat{table}
\usepackage[caption=false]{subfig}
\usepackage{showlabels}
\RequirePackage[colorlinks,citecolor=blue,urlcolor=magenta,linkcolor=blue]{hyperref}
\input epsf
\usepackage[labelsep=none,labelformat=empty]{caption}
\def\gr{general relativity}
\def\RN{Reissner-Nordstr\"{o}m }
\def\KN{Kerr-Newmann }

\labelformat{section}{Section #1} 
\labelformat{subsection}{Section #1} 
\labelformat{subsubsection}{Section #1}
\labelformat{subsubsubsection}{Section #1}
\labelformat{equation}{Eq.~(#1)} 
\labelformat{figure}{Fig.~#1} 
\labelformat{subfigure}{Fig.~\thefigure#1} 
\labelformat{table}{Table~#1} 
\labelformat{appendix}{Appendix #1}
\title{Decoding signatures of extra dimensions and estimating spin of quasars from the  continuum spectrum}
\author{Indrani Banerjee\footnote{tpib@iacs.res.in}~$^{1}$, Sumanta Chakraborty\footnote{sumantac.physics@gmail.com}~$^{1}$ and Soumitra SenGupta\footnote{tpssg@iacs.res.in}~$^{1}$\\
{\small{$^{1}$School of Physical Sciences, Indian Association for the Cultivation of Science, Kolkata-700032, India}}}
\begin{document}
  
\maketitle
\begin{abstract}
Continuum spectrum emitted by the accretion disk around quasars hold a wealth of information regarding the strong gravitational field produced by the massive central object. Such strong gravity regime is often expected to exhibit deviations from \gr$~$(GR) which may manifest through the presence of extra dimensions. Higher dimensions, which serve as the corner stone for string theory and M-theory can act as promising alternatives to dark matter and dark energy with interesting implications in inflationary cosmology, gravitational waves and collider physics. Therefore it is instructive to investigate the effect of more than four spacetime dimensions on the black hole continuum spectrum which provide an effective astrophysical probe to the strong gravity regime. To explore such a scenario, we compute the optical luminosity emitted by a thin accretion disk around a rotating supermassive black hole albeit in the presence of extra dimensions. The background metric resembles the \KN spacetime in GR where the tidal charge parameter inherited from extra dimensions can also assume negative signature. The theoretical luminosity computed in such a background is contrasted with optical observations of eighty quasars. The difference between the theoretical and observed luminosity for these quasars is used to infer the most favoured choice of the rotation parameter for each quasar and the tidal charge parameter. This has been achieved by minimizing/maximizing several error estimators, e.g., $\chi^{2}$, Nash-Sutcliffe efficiency, index of agreement etc. Intriguingly, all of them favour a \emph{negative} value for the tidal charge parameter, a characteristic signature of extra dimensions. Thus accretion disk does provide a significant possibility of exploring the existence of extra dimensions through its close correspondence with the strong gravity regime. 
\end{abstract}
\section{Introduction}\label{Accretion_Intro}

General Relativity (GR) is the most successful theory, so far, in explaining the gravitational interaction at various length scales. The success of GR stems from its remarkable theoretical predictions and their consistency with observations \cite{Will:2005yc,Will:1993ns,Will:2005va,Berti:2015itd}. The most recent in the list being the detection of gravitational waves from colliding black holes and neutron stars \cite{Abbott:2017vtc,TheLIGOScientific:2016pea,Abbott:2016nmj,TheLIGOScientific:2016src,Abbott:2016blz}. The associated field equations, known as Einstein's equations, govern the dynamics of an apple falling to the ground as well as the evolution of the universe as a whole \cite{Lemaitre:1931zz,Robertson:1935zz,Robertson:1936zza,Robertson:1936zz}. Despite its brilliant success over such a large length scale, there are a few regimes where the theory breaks down. The most notable among them are the black hole and cosmological singularities, where the theory loses its predictive power \cite{Penrose:1964wq,Hawking:1976ra,Wald,Christodoulou:1991yfa}. This is tantamount to saying that at very small length scales (possibly Planck length or less) GR must receive significant corrections from a more complete theory of gravitation that incorporates its quantum character \cite{Rovelli:1996dv,Dowker:2005tz,Ashtekar:2006rx,Kothawala:2013maa}. The lack of an undisputed understanding of the nature of strong gravity coupled with the inadequacy of GR to address the pressing issues like dark matter \cite{Milgrom:1983pn,Bekenstein:1984tv,Milgrom:2003ui} and dark energy \cite{Clifton:2011jh,Perlmutter:1998np,Riess:1998cb}, makes the quest for a more complete theory of gravity increasingly compelling. This has led to a proliferation of alternate gravity models which can be strong contenders of GR.

Such alternate gravity models should not only pass the classic tests of GR but also offer explanations to the unresolved issues like dark matter and/or dark energy. Consequently, there have been several approaches to modify the structure of the gravitational action. Among the various proposals the most prominent ones include, e.g., $f(R)$ gravity \cite{Nojiri:2010wj,Nojiri:2003ft,Nojiri:2006gh,Capozziello:2006dj,Bahamonde:2016wmz,Chakraborty:2016ydo}, Lovelock theories \cite{Lanczos:1932zz,Lanczos:1938sf,Lovelock:1971yv,Padmanabhan:2013xyr,Dadhich:2015ivt,Chakraborty:2014joa,Chakraborty:2015wma}, scalar-tensor theories/Horndeski models \cite{Horndeski:1974wa,Sotiriou:2013qea,Babichev:2016rlq,Brans:1961sx,VanAcoleyen:2011mj,Parattu:2016trq,Charmousis:2015txa,Bhattacharya:2016naa,Antoniou:2017acq,Bakopoulos:2018nui} and theories with extra spatial dimensions \cite{Shiromizu:1999wj,Dadhich:2000am,Harko:2004ui,Carames:2012gr,Kobayashi:2006jw,Shiromizu:2002qr,Haghani:2012zq,Borzou:2009gn,Chakraborty:2014xla,Chakraborty:2015bja,Chakraborty:2015taq,Chakraborty:2016gpg} (see also \cite{Nojiri:2017ncd}). 

In this work we explore modifications in GR due to extra dimensions since it offers one of the simplest and minimal alteration to the Einstein-Hilbert action. Extra dimensions, which serve as the bedrock for string theory and M-theory \cite{Horava:1995qa,Horava:1996ma,Polchinski:1998rq,Polchinski:1998rr}, arise naturally in the quest for unification of all the known forces. The string-inspired brane-world models \cite{Randall:1999vf,Randall:1999ee,Garriga:1999yh,Csaki:1999mp} which are designed to provide a resolution to the fine-tuning problem in particle physics \cite{Antoniadis:1990ew,ArkaniHamed:1998rs,Antoniadis:1998ig,Randall:1999ee,Csaki:1999mp,Garriga:1999yh,Goldberger:1999uk,Banerjee:2017jyk} also owe their origin to extra dimensions. These models assume that the observable Universe comprising of the Standard Model particles and fields are confined to a 3-brane while gravity pervades the bulk \cite{Dadhich:2000am,Harko:2004ui,Carames:2012gr,Kobayashi:2006jw,Shiromizu:2002qr,Haghani:2012zq,Borzou:2009gn,Chakraborty:2014xla,Chakraborty:2015bja,Chakraborty:2015taq,Chakraborty:2016gpg,ArkaniHamed:1998rs,Shiromizu:1999wj,Randall:1999vf}. Motivated by the success of the brane world models in several theoretical contexts, it is important to subject them to various observational tests, wherever there is a scope. The most widely used observational probe for extra dimensions come from particle physics experiments \cite{Dimopoulos:2001hw,Davoudiasl:1999jd,Davoudiasl:1999tf,Davoudiasl:2002fq,Chung:2000rg,ArkaniHamed:1998nn,Banks:1999gd}. In addition, they have interesting implications in inflationary cosmology \cite{Lukas:1998qs,Lukas:1999yn,ArkaniHamed:1999gq,Dienes:1998hx,Mazumdar:1999tk,Mazumdar:2000sw,Mazumdar:2003vg,Chakraborty:2013ipa,Banerjee:2017lxi,Paul:2018kdq,Banerjee:2018kcz} and often offer exciting surrogates to the elusive dark energy \cite{Shiromizu:1999wj,Dadhich:2000am,Harko:2004ui,Carames:2012gr,Kobayashi:2006jw,Shiromizu:2002qr,Haghani:2012zq,Borzou:2009gn,Chakraborty:2014xla,Chakraborty:2015bja,Chakraborty:2015taq,Chakraborty:2016gpg} and dark matter \cite{Pal:2004ii,Capozziello:2006uv,Boehmer:2007az,Harko:2007yq,Chakraborty:2015zxc}. In the context of gravitational waves they modify the quasi-normal modes emanating from the perturbed black holes with signatures distinct from general relativity \cite{Berti:2009kk,Kanti:2005xa,Konoplya:2011qq,Toshmatov:2016bsb,Andriot:2017oaz,Chakraborty:2017qve}. The presence of extra dimensions also affect the equation of state of a neutron star which in turn alter the tidal Love numbers measuring its deformability \cite{Hinderer:2007mb,Flanagan:2007ix,DelPozzo:2013ala,Yagi:2016ejg,Cardoso:2017cfl}. This can be used to impose tight constraints on the extra dimensional parameters, using for example, the GW170817 event \cite{Visinelli:2017bny,Abbott:2017ntl,TheLIGOScientific:2017qsa,Chakravarti:2018vlt}.

In this work, we aim to discern the footprints of extra dimensions in black hole accretion. We consider a single brane encompassing the visible universe, embedded in a five dimensional bulk. As a consequence, the non-local effects of the bulk Weyl tensor acts as a source to four dimensional gravity, even in the absence of any matter-energy on the brane \cite{Shiromizu:1999wj,Shiromizu:2002qr,Kanno:2002ia,Harko:2004ui,Chakraborty:2015bja}. A certain class of vacuum solutions of the modified field equations resemble the \RN metric in GR if the background spacetime is static and spherically symmetric \cite{Dadhich:2000am,Harko:2004ui} or the \KN solution in case the metric is stationary and axi-symmetric \cite{Aliev:2005bi}. However, unlike GR, the tidal charge parameter in the aforesaid spacetimes owes its origin to extra dimensions and hence can also assume negative values which turns out to be a distinguishing signature of higher dimensons \cite{Schee:2008kz}. 

Continuum spectrum emitted by the accretion disk around quasars encapsulate enormous information regarding the nature of the background spacetime and hence serve as promising probes to the strong gravity regime around quasars. The associated accretion flow is approximated by the geometrically thin and optically thick accretion disk discussed in \cite{Shakura:1972te,Thorne:1974ve,Page:1974he,Novikov_Thorne_1973}. Since the thin accretion disk around supermassive black holes emit chiefly in optical frequencies \cite{2002apa..book.....F} we compute the theoretical estimates of optical luminosity for a sample of eighty Palomar Green quasars \cite{Schmidt:1983hr,Davis:2010uq} assuming the background metric to mimic the aforesaid \KN spacetime. Therefore, the theoretical luminosity depends not only on the tidal charge parameter inherited from extra dimensions but also on the rotation of the quasars.  

This when compared with the corresponding optical observations of quasars provide a cue not only on the signature of the tidal charge parameter but also an estimate on the rotation parameters of the quasars. Previously, we explored the effect of the tidal charge parameter alone on the theoretical luminosity and reported that quasar optical data favor a negative charge parameter \cite{Banerjee:2017hzw}. Since quasars are intrinsically rotating, an extension of \cite{Banerjee:2017hzw} is essential by incorporating the spin of the quasars in our analysis. This is the primary goal of this work which in turn provides a greater reliance on our previous findings. Our conclusions are based on several tests of goodness-of-fit, e.g., chi-squared, Nash-Sutcliffe efficiency and index of agreement which renders a more quantitative appreciation of our results. 

The paper is organized as follows: In \ref{Accretion_Alt} we describe the rotating black hole solution and the brane world model we are considering. The basic steps of calculating the flux and hence the luminosity from the accretion disk in such a black hole spacetime have been explored in \ref{acc_axisym}. Subsequently in \ref{Accretion_Obs}, a comparison between the theoretical and observed luminosity of eighty quasars is performed in terms of several error estimators. Finally, we conclude with a discussion of our results and some scope for future work in \ref{Accretion_Conc}.       

\textit{Notations and Conventions:} Throughout this paper, the Greek indices denote the four dimensional spacetime and we will work in geometrized unit with $G=1=c$ and the metric convention will be mostly positive. 
\section{Rotating black hole in the brane world gravity}\label{Accretion_Alt}
 
In this section we will outline the derivation of the vacuum solutions of the modified gravitational field equations as perceived by a four dimensional observer in the presence of higher dimension. There are several ways of achieving the same, possibly by integrating out the bulk degrees of freedom or by projecting the bulk gravitational field equations on the brane hypersurface. In this work, we will adopt the second approach, since the other avenues require a detailed knowledge of the bulk geometry which a priori is absent to a brane observer. Therefore, the approach taken by us is much more generic. In this particular scenario, one projects the bulk Einstein's equations on the brane hypersurface and through the use of Gauss and Codazzi equations it is possible to arrive at the effective gravitational field equations on the brane \cite{Shiromizu:1999wj,Harko:2004ui,Chakraborty:2014xla,Chakraborty:2015bja}. This involves the four-dimensional Einstein tensor $G_{\mu \nu}$ as well as the bulk Weyl tensor, projected onto the brane hypersurface. There are of course additional contributions from the matter sector, which however vanish in the context of a vacuum brane. The additional piece generated from the bulk Weyl tensor (also known as the electric part of the Weyl tensor), by virtue its inherent symmetries is traceless and hence mimics the energy-momentum tensor of a Maxwell field \cite{Aliev:2005bi}. The static and spherically symmetric spacetime in this scenario looks like the Reissner-Nordstr\"{o}m solution, where the charge term can be both positive and negative. The negative value of the charge term provides a characteristic signature of these extra dimensional models. The above static and spherical symmetric brane spacetime has already been investigated in \cite{Dadhich:2000am,Banerjee:2017hzw}, following which in this work we want to explore the stationary and axisymmetric counterpart of the same \cite{Aliev:2005bi}, which is observationally more relevant. 

In the context of any stationary and axi-symmetric spacetime there exists two Killing vectors $(\partial/\partial t)^{a}$ and $(\partial/\partial \phi)^{a}$. Further, absence of any brane matter enables one to express the line element in terms of five functions of radial and angular coordinate. In this particular context, the solution looks like the Kerr-Newman spacetime with the associated line element,
\begin{align}\label{2-2}
ds^{2}&=-\bigg(1-\frac{2Mr-4M^{2}q}{\rho^{2}}\bigg)dt^2-\frac{2a\sin^2\theta(2Mr-4M^{2}q)}{\rho^{2}} dt d\phi 
+ \frac{\rho^{2}}{\Delta}dr^2 +\rho^{2} d\theta^2 
\nonumber
\\
&\qquad \qquad+\bigg\{r^2 + a^2 +\frac{a^2 \sin^2\theta\left(2Mr-4M^{2}q\right)}{\rho^{2}}\bigg\}\sin^2\theta d\phi^2~,
\end{align}
where we have defined, $\rho^{2}=r^2+a^2 cos^2\theta$ and $\Delta=r^2 -2Mr+a^2+4M^{2}q$. \\
Note that the charge parameter $q$ in \ref{2-2} is inherited from the presence of higher dimensions and hence can assume both positive and negative values. For positive $q$, \ref{2-2} appears similar to a \KN black hole with an event horizon and a Cauchy horizon, while the case with negative $q$ has no analogue in \gr\ and thus provides a true signature of the additional spatial dimensions \cite{Aliev:2005bi}.

In what follows we will analyze the motion of particles in this black hole spacetime, acting as the basic constituents of the accretion disk. In a simplified scenario, which we will consider in this work, the accretion flow will be restricted to the equatorial plane (i.e., $\theta=\pi/2$) with the accreting particles following nearly circular geodesics. This implies that the resultant accretion disk is geometrically thin, i.e., $\{h(r)/r\}\ll 1$, where $h(r)$ is the disk height at a radial distance $r$ from the black hole. Since the flow is along the equatorial plane the metric coefficients are only functions of the radial coordinate $r$. Since the motion is circular and on the equatorial plane, the azimuthal velocity $v_{\phi}$ is the dominant one. While for accretion to take place the accreting matter should also have a minimal radial velocity $v_{r}$ and negligible vertical velocity $v_{z}$, i.e., $v_{z} \ll v_{r} \ll v_{\phi}$, such that one ends up with a thin accretion disk with no outflows. The small radial velocity is gained due to viscous dissipation which facilitates loss of angular momentum and results into gradual in-spiral and ultimate fall of matter into the black hole.
\section{Luminosity from the accretion disk around a rotating brane world black hole}\label{acc_axisym}

In this section we will discuss various properties of electromagnetic emission from the accretion disk around a rotating brane world black hole and shall use them to quantify the luminosity from the accretion disk. For generality we will present some of the results for a general stationary and axi-symmetric spacetime before specializing to the particular case presented in \ref{2-2}. As illustrated above, such a general stationary and axi-symmetric background is encoded by five functions, namely, $g_{tt}$, $g_{t\phi}$, $g_{\phi \phi}$, $g_{rr}$ and finally $g_{\theta \theta}$. However for motion in the equatorial plane, the $g_{\theta \theta}$ part does not contribute. Since any stationary and axi-symmetric spacetime has two Killing vectors, viz., $(\partial/\partial t)^{a}$ and $(\partial/\partial \phi)^{a}$, there are two conserved quantities, namely, the specific energy $E$ and the specific angular momentum 
$L$. For massive test particles moving in circular geodesics the specific energy can be expressed in terms of the metric elements as,
\begin{align}\label{3-2}
E=\frac{-g_{tt}-\Omega g_{t\phi}}{\sqrt{-g_{tt}-2\Omega g_{t\phi}-\Omega^2 g_{\phi\phi}}}~.
\end{align}
On the other hand the specific angular momentum takes the form,
\begin{align}\label{3-3}
L=\frac{\Omega g_{\phi\phi}+g_{t\phi}}{\sqrt{-g_{tt}-2\Omega g_{t\phi}-\Omega^2 g_{\phi\phi}}}~,
\end{align}
where, $\Omega=(d\phi/dt)$ is the angular velocity associated with the circular trajectory of the massive particle. Using the geodesic equation it is possible to express the angular velocity as well in terms of metric coefficients,
\begin{align}\label{3-4}
\Omega=\frac{d\phi}{dt}=\frac{-g_{t\phi,r}\pm \sqrt{\left\lbrace-g_{t\phi,r}\right\rbrace^2-\left\lbrace g_{\phi\phi,r}\right\rbrace \left\lbrace g_{tt,r}\right\rbrace}}{g_{\phi\phi,r}}~,
\end{align}
where the positive (negative) sign in \ref{3-4} corresponds to prograde (retrograde) orbits respectively. Since both $g_{tt}$ and $g_{t\phi}$ are negative, the quantity under the square root turns out to be positive, leading to non-trivial expressions for $\Omega$. These three quantities, namely angular velocity, specific energy and specific angular momentum associated with the circular motion on the equatorial plane are the most important quantities one requires in order to compute the flux and hence the luminosity emanating from the accretion disk.  

To proceed further, we need to write down the energy momentum tensor associated with the accretion flow, which may be treated as a fluid. Using the four velocity of the accreting particles moving in circular geodesics, one can express the energy momentum tensor as,
\begin{align}\label{3-5}
T^{\mu}_ {\nu}= \rho_{0}\left(1+\Pi\right)u^{\mu}u_{\nu}+t^{\mu}_{\nu}+u^{\mu}q_{\nu}+q^{\mu}u_{\nu}~.
\end{align}
Here $u^{\alpha}$ stands for the four velocity of the particles moving with the accretion flow and $\rho _{0}$ is the energy density measured by the accreting matter if treated as a perfect fluid. On the other hand, the quantity $\Pi$ represents contribution to the energy density due to dissipation, i.e., departure of the accreting material from a perfect fluid and is referred to as the specific internal energy of the system. Finally $t^{\alpha \beta}$ is the stress-tensor measured in the local rest frame of the accreting fluid and $q^{\alpha}$ is the energy flux relative to the local inertial frame. As a consequence, $t_{\alpha \beta}u^{\alpha}=0=q_{\alpha}u^{\alpha}$. The stability of the accretion disk and the fact that the accreting particles follow nearly circular geodesics imply that the gravitational pull of the central black hole supercedes the forces due to radial pressure gradients. Therefore the specific internal energy of the accreting fluid, i.e., $\Pi$ can be neglected compared to the rest energy of the fluid, such that $\Pi\ll 1$. This translates to the fact that although one needs to consider general relativistic effects due to the presence of the black hole, the special relativistic corrections to the local hydrodynamic, thermodynamic and radiative properties of the fluid can be safely ignored.
Consequently, the amount of energy generated due to the viscous dissipation is completely radiated away and hence no heat is trapped with the accreting matter. Therefore, only the z-component of the energy flux vector $q^\alpha$ has a non-zero contribution. For a more detailed exposition on the assumptions associated with the thin accretion disk model we refer the reader to \cite{Novikov_Thorne_1973,Page:1974he}.

\subsection{The Conservation Laws}

Having discussed the basic overview of the theoretical properties of the accretion disk, we now present the computation of flux and hence luminosity arising from the accretion into the central black hole. This will be achieved by introducing three conservation laws, namely the conservation of mass, angular momentum and energy. In what follows we will briefly discuss all of these conservation laws and finally we will present the expression for flux from the accretion disk. 
\begin{itemize}

\item {\bf Conservation of mass of the accreting fluid:} By conservation of mass we mean, if we consider a small volume, then the rate of change of mass within that volume compensates for the total mass flowing out of that volume. In the context of accretion one can safely assume that over the time scales of interest, the matter is accreted at a nearly constant rate $\dot{M}_{0}$, such that in a small time interval $\Delta t$, the total amount of mass influx at a fixed radius $r$ corresponds to $\dot{M_0}\Delta t$. This should be equal to the mass flowing out of $r=\textrm{constant}$ surface and hence to the integral of the radial mass current $\rho_0 u^r$ over the $r=\textrm{constant}$ hypersurface. Thus the integration measure will correspond to $\sqrt{-g_{tt}g_{\phi \phi}}dt d\phi dz$, where instead of $\theta$ we are using $z$ as the vertical coordinate. Since over the time scale of interest the accretion disk+black hole system can be considered to be stationary as well as axi-symmetric, we will work with time-averaged quantities. Thus the three dimensional integral over $\rho_{0}u^{r}$ can be converted to an integral over $z$ of $\langle \rho_{0}\rangle$, which yields,
\begin{align}\label{3-7}
\dot{M}_{0}=-2\pi \sqrt{-g}u^{r} \Sigma;\qquad \Sigma =\int \left\langle \rho_0 \right\rangle dz
\end{align}
where $g$ represents determinant of the axi-symmetric metric, which in our particular case of brane world black hole corresponds to \ref{2-2}. Further the quantity $\Sigma$ defined above depicts the average surface density of matter flowing into the black hole. Thus conservation of mass relates the accretion rate $\dot{M}_{0}$ with the average surface density. We will now consider the other two conservation relations.

\item {\bf Conservation of angular momentum of the accreting fluid:} The conservation of angular momentum is best described by defining a current starting from the energy-momentum tensor $T_{\mu \nu}$ defined in \ref{3-5}, such that $J^{\mu}_{\phi}=T^{\mu}_{\nu}(\partial/\partial \phi)^{\nu}$. From the conservation of energy-momentum tensor and Killing equation for $(\partial/\partial \phi)^{\nu}$, it is clear that $J_{\phi}^{\mu}$ is conserved, i.e., $\nabla_{\mu}J_{\phi}^{\mu}=0$. Using the expression for $T^{\mu}_{\nu}$ from \ref{3-5}, the conservation of $J^{\mu}_{\phi}$ translates into the following differential equation, 
\begin{align}\label{3-10}
\rho_0\left\lbrace u^\alpha \nabla_\alpha u_\phi\right\rbrace +\frac{1}{\sqrt{-g}}\partial_r (\sqrt{-g} t^r_\phi) + u_\phi \partial_z q^z=0 
\end{align}
where we have assumed geodesic flow along the equatorial plane of the black hole, have neglected contribution from the internal energy term $\Pi$ and have used the orthogonality conditions ${u^{\nu} t^{\mu}{_\nu}=0=u^\mu q_\mu}$. Subsequent integration over time and $z$ leads to the following quantity, $\langle q^{z}(r,h)\rangle -\langle q^{z}(r,-h)\rangle$ from the last term of \ref{3-10} as well as integration of $\langle t^r_\phi \rangle$ over the height of the disk from the second term. The first term is related to $\Sigma$ and hence to the accretion rate $\dot{M}_{0}$, thanks to mass conservation as presented in \ref{3-7}. Therefore, the conservation of angular momentum assumes the form, 
\begin{align}\label{3-13}
\frac{\partial}{\partial r}\left[\dot{M_0}L-2\pi \sqrt{-g}W^r_\phi \right]=4\pi \sqrt{-g} F L~,
\end{align}
where $F$ corresponds to the flux of radiation generated by the accretion process and is defined as, $F\equiv \langle q^{z}(r,h)\rangle=-\langle q^{z}(r,-h)\rangle$ and $W^{r}_{\phi}$ denotes the height and time averaged stress tensor in the local rest frame of the accreting particles. Thus the conservation of angular momentum can be casted as the differential equation presented above, involving flux, accretion rate and the average local stress tensor. 

\item {\bf Conservation of energy of the accreting fluid:} Finally, we consider conservation of energy of the accreting fluid. As in the case of angular momentum, for energy as well, one may define the current $J^{\mu}_{t}=T^{\mu}_{\nu}(\frac{\partial}{\partial t})^{\nu}$. Conservation of energy-momentum tensor and Killing equation satisfied by $(\frac{\partial}{\partial t})^{\mu}$ ensures that $J^{\mu}_{t}$ is also conserved. Hence expanding out the relation $\nabla_{\mu}J^{\mu}_{t}=0$, we can rewrite the conservation of energy as the following differential equation,
\begin{align}\label{3-11}
\rho_0\left\lbrace u^\alpha \nabla_\alpha u_t\right\rbrace +\frac{1}{\sqrt{-g}}\partial_r (\sqrt{-g} t^r_t) + u_t\partial_z q^z=0
\end{align}
Integration over $t$ and $z$ coordinates yields expectation values of various quantities averaged along the vertical direction. As for the third term in the above expression this yields $\langle q^{z}(r,h)\rangle -\langle q^{z}(r,-h)\rangle$, which equals to $2F$ by virtue of the definition of the flux from accretion disk. Along identical lines we obtain, time and height average of $t^{r}_{t}$ to be $W^{r}_{t}$. The first term in \ref{3-11} can be related to the surface density $\Sigma$ and hence to the accretion rate $\dot{M}_{0}$ through the use of \ref{3-7}. It turns out that the quantity $W^{r}_{\phi}$ and $W^{r}_{t}$ are not independent, rather they are connected by a certain relation. This can be derived by invoking the orthogonality relation, $u^{\nu} t^{\mu}{_\nu}=0$ which further implies $u^{\nu} W^{\mu}{_\nu}=0$ as well. Hence one can write,
\begin{align}\label{3-15}
W^r_t=-\frac{u^\phi}{u^t}W^r_\phi =-\Omega W^r_\phi
\end{align}
where $\Omega$ is the angular velocity of the accreting particles. Using this relation, we can express the condition for the conservation of energy as,
\begin{align}\label{3-16}
\frac{\partial}{\partial r}\left[\dot{M_0}E-2\pi \sqrt{-g}\Omega W^r_\phi \right]=4\pi \sqrt{-g} F E 
\end{align}
In this expression as well, $F$ stands for the flux from the accretion disk and $W^{r}_{\phi}$ is the time and height averaged local stress energy tensor. In the remaining discussion, we will eliminate the term $W^{r}_{\phi}$ and shall express the flux in terms of energy, angular momentum and angular velocity of the accreting particles alone.

\end{itemize}

\subsection{Flux from the accretion disk}

With all the conservation relations at hand, in this section, we proceed to compute the flux of electromagnetic radiation emitted from the accretion disk. For this purpose we define a normalized flux $f$ such that, $f=4\pi\sqrt{-g}(F/\dot{M}_{0})$ as well as normalized local stress tensor, $w=2\pi\sqrt{-g}(W^{r}_{\phi}/\dot{M}_{0})$. In terms of these two quantities we may rewrite the angular momentum and energy conservation relations as presented in \ref{3-13} and \ref{3-16} in the following form,
\begin{align}\label{3-17}
\partial _{r}\left(L-w\right)=fL;
\qquad
\partial _{r}\left(E-\Omega w\right)=fE
\end{align}
Further, taking into account the geodesic equation $u^\mu\nabla_\mu u^\nu=0$ satisfied by the accreting particles and the normalization condition of the $4-$velocity, i.e., $u^\mu u_\mu=-1$ one can relate the energy of the particle with its angular momentum through the angular velocity $\Omega$ as, $\partial _{r}E=\Omega \partial _{r}L$. Use of this relation in \ref{3-17} allows one to solve the simultaneous differential equations for both the normalized flux $f$ and the normalized local stress tensor $w$, yielding
\begin{align}\label{3-21}
f=-\frac{\partial _{r}\Omega}{(E-\Omega L)^2}\int dr (E-\Omega L)\partial _{r}L+ \textrm{constant};
\qquad
w=-\frac{E-\Omega L}{\partial _{r}\Omega}f
\end{align}
The constant of integration appearing in the expression for flux, presented in \ref{3-21} is fixed by assuming that the accretion disk truncates at the marginally stable circular orbit $r_{\rm ms}$, i.e., the accreting fluid ceases to have any azimuthal velocity after crossing $r_{ms}$. Consequently, the shear stress $W^r_\phi$, or equivalently $w$ vanishes at $r_{\rm ms}$. In such a scenario, the expression for flux, as in \ref{3-21} can be expressed as,
\begin{align}\label{3-23}
\tilde{f}=-\frac{\partial _{r}\Omega}{(E-\Omega L)^2}\int^{r}_{r_{\rm ms}}d\bar{r}(E-\Omega L)\partial _{\bar{r}}L
\end{align}
The above expression can be casted in several alternative forms by using integration by parts and the result $\partial _{r}E=\Omega \partial_{r}L$. However, for this work we will concentrate with the above expression for $f$, from which the flux $F$ emanating from the accretion disk at a radial distance $r$ is given by,
\begin{align}\label{3-25}
F = \frac{\dot{M}_{0}}{4\pi\sqrt{-g}}f=-\frac{\dot{M}_{0}\partial _{r}\Omega}{4\pi\sqrt{-g}(E-\Omega L)^2}\int^{r}_{r_{\rm ms}}d\bar{r}(E-\Omega L)\partial _{\bar{r}}L
\end{align}
This is the main observable we will consider in this work. Note that the photon produced from the accretion process will undergo repeated collisions with the accreting fluid, leading to a thermal equilibrium between accreting matter and emanating radiation. This renders the disk to be geometrically thin but optically thick and enables it to emit a blackbody radiation, locally.

Due to the blackbody nature of the emitted radiation, the temperature profile of the disk is obtained from the Stefan-Boltzmann law, such that $F_{\rm act}(r)=\sigma T(r)^{4}$. Here $\sigma$ is the Stefan-Boltzmann constant and $F_{\rm act}(r)=F(r)c^6/(G^2 M^2)$ is obtained from \ref{3-25} by restoring all the relevant fundamental constants. Thus, the disk emits a Planck spectrum at every radial distance $r$ with a peak temperature $T(r)$. However for convenience one often introduces a dimensionless quantity $x$, defined as $x=r/R_{\rm g}$, where $R_{\rm g}=GM/c^{2}$ is the gravitational radius. Hence the luminosity $L_\nu$ emitted by the disk at a frequency $\nu$ is obtained by integrating the Planck function $B_\nu(T(x))$ over the disk surface, yielding,
\begin{align}\label{3-26}
L_{\nu}=8\pi^2 R_{\rm g}^2\cos i  \int_{x_{\rm ms}}^{x_{\rm out}}\sqrt{g_{rr}} B_{\nu}(T)x dx;
\qquad
B_\nu (T)&=\frac{2h\nu^3/c^2}{{\rm exp}\left(\frac{h\nu}{z_g kT}\right)-1}
\end{align}
In the above expression $i$ stands for the inclination angle between the normal to the disk and the line of sight, while the gravitational redshift factor $z_g$ in \ref{3-26} is given by,
\begin{align}\label{3-27}
z_g=E\frac{\sqrt{-g_{tt}-2\Omega g_{t\phi}-\Omega^2 g_{\phi\phi}}}{E-\Omega L}
\end{align}
which relates the change in the frequency suffered by the photon while travelling from the emitting material to the observer \cite{Ayzenberg:2017ufk}. Thus we have finally arrived at the theoretical expression for the luminosity originating from the accretion disk. It depends on the mass of the central black hole, the accretion rate, the inclination angle and the metric parameters through the energy, angular momentum, angular velocity and the radius of marginally stable circular orbit. In the subsequent section we discuss expressions of these quantities in the brane world black hole spacetime under consideration. This will provide us the desired handle to understand the effect of extra dimensions on the continuum spectrum from quasars.  
 
\subsection{Specializing to brane world black hole} 

In this section we will explicitly compute the angular velocity, the specific energy and the specific angular momentum associated with the rotating brane world black hole spacetime, whose line element is given by \ref{2-2}. For this purpose it is important to first write down the relevant metric elements on the equatorial plane, 
\begin{align}\label{3-30}
g_{tt}=-\left(1-\frac{2x-4q}{x^2}\right);~~~~~~g_{t\phi}=-a_{*}\frac{(2x-4q)}{x^2};~~~~~~g_{\phi\phi}=x^2+a_{*}^2+a_{*}^2\frac{(2x-4q)}{x^2};
\end{align}
Here we have introduced the dimensionless radial coordinate $x=r/M$, as well as dimensionless rotation parameter $a_{*}=a/M$ and dimensionless charge parameter $q$. In terms of these quantities, one can use \ref{3-4} along with \ref{3-30} in order to derive the angular velocity $\Omega$ in terms of the metric elements, which reads,
\begin{align}\label{3-28}
\Omega=\frac{\sqrt{x-4q}}{M\left(x^2+a_{*}\sqrt{x-4q}\right)}
\end{align}
Along identical lines one can express the specific energy $E$ and specific angular momentum $L$ in terms of various parameters appearing in the black hole metric, which by the use of \ref{3-2} and  \ref{3-3} takes the following form,
\begin{align}
E&=\frac{1-\frac{2}{x}+\frac{q}{x^{2}}+\frac{a_{*}}{x}\sqrt{1-\frac{q}{x^{2}}}}{\sqrt{1-\frac{3}{x}+2\frac{q}{x^{2}}+2\frac{a_{*}}{x}\left(1-\frac{q}{x^{2}}\right)}}
\label{3-29}
\\
L&=M\sqrt{x}\frac{\left(1+\frac{a_{*}^{2}}{x^{2}}\right)\sqrt{1-\frac{q}{x}}-2\frac{a_{*}}{x\sqrt{x}}\left(1-\frac{q}{2x} \right)}{\sqrt{1-\frac{3}{x}+2\frac{q}{x^{2}}+2\frac{a_{*}}{x}\left(1-\frac{q}{x^{2}}\right)}}
\label{3-31}
\end{align}
Thus with the above expressions for angular velocity, specific energy and angular momentum expressed in terms of specific rotation parameter $a_{*}$ as well as the tidal charge parameter $q$ we have most of the ingredients required to estimate the luminosity in this context. The only remaining quantity correspond to the radius of the marginally stable orbit, i.e., $r_{\rm ms}$. This is accomplished by finding the effective potential $V_{\rm eff}$ in which the individual particles constituting the accreting fluid moves. One can write down an expression for the effective potential in a stationary and axi-symmetric spacetime as,
\begin{align}\label{3-32}
V_{\rm eff}(r)=\frac{E^2g_{\phi\phi}+2ELg_{t\phi}+L^2g_{tt}}{g_{t\phi}^2-g_{tt}g_{\phi\phi}}-1
\end{align}
where $g_{tt}$, $g_{t\phi}$ and $g_{\phi\phi}$ are the metric elements given in \ref{3-30}, while $E$ and $L$ are the specific energy and specific angular momentum presented in \ref{3-29} and \ref{3-31} respectively. The marginally stable circular orbit corresponds to the inflection point of the effective potential, which is obtained by solving for $V_{\rm eff}=\partial _{r}V_{\rm eff}=0=\partial _{r}^{2}V_{\rm eff}$. For the brane world black hole spacetime under consideration, the dimensionless marginally stable circular orbit, $x_{\rm ms}$ is obtained by solving for the lowest real root of \cite{Blaschke:2016uyo},  
\begin{align}\label{3-33}
x\left(6x-x^2-36q+3a_{*}^2\right)+16q(4q-a_{*}^2)-8a_{*}(x-4q)^{3/2}=0
\end{align}
As evident the radius of marginally stable circular orbit $x_{\rm ms}$ will be a function of tidal charge parameter $q$ and specific rotation parameter $a_{*}$. One can indeed verify that, for $q=0=a_{*}$, $x_{\rm ms}=6$, while for $q=0$ and $a_{*}=\pm 1$, $x_{\rm ms}=1,9$ respectively. This explicitly demonstrates that we get back the Schwarzschild and the Kerr geometry correctly under appropriate limits. In order to derive the radial dependence of the flux, one needs to compute the integral presented in \ref{3-25}, calculated with $r_{\rm ms}$ as the lower limit. This can be accomplished by using the metric elements of the rotating black hole spacetime and solving for $r_{\rm ms}(a,q)$ from \ref{3-33}. Once the flux is known, the luminosity can be computed by using \ref{3-26}.

\begin{figure}[t!]
\centering
\includegraphics[scale=0.75]{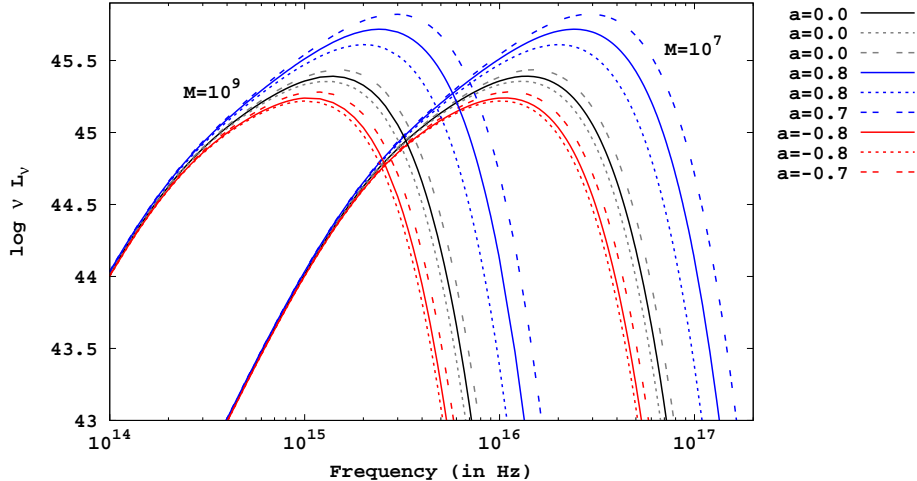}
\caption{Figure 1: The above figure depicts variation of the theoretically derived  luminosity from the accretion disk with frequency for two different masses of the supermassive black holes. For both the masses, the solid lines represent $q=0$, dotted lines represent $q=-0.4$, and dashed lines represent $q=0.4$. The solid black line corresponds to the Schwarzschild scenario. Note that for a fixed $q$, increase in $a$ corresponds to increased luminosity. Similarly for a fixed $a$, a positive value of $q$ leads to greater peak luminosity than the corresponding $q=0$ or negative $q$ situations. The accretion rate assumed is $1 M_{\odot}\textrm{yr}^{-1}$ and $\cos i$ is taken to be $0.8$. This result holds for supermassive black holes with masses $M=10^7 M_\odot$ and $M=10^9 M_\odot$ although the peak frequency is higher for a lower mass quasar. This is because the peak temperature $T$ for a multi-color black body spectrum scales as $T \propto M^{-1/4}$ \cite{Frank:2002}. See text for more discussions.}
\label{Fig_01}
\end{figure}

The above discussion elucidates that the theoretically estimated luminosity from the accretion disk $L_{\rm opt}$ depends on the mass $M$ of the black hole, the accretion rate $\dot{M_0}$, the inclination angle $i$ and the nature of the background spacetime. For brane world black holes, the background spacetime is sensitive to two parameters, namely, the specific rotation parameter $a_{*}$ of the black hole and the tidal charge parameter $q$. The variation of theoretical luminosity with frequency for known mass and accretion rate is demonstrated in \ref{Fig_01} which illustrates that both $q$ and $a_{*}$ affect the theoretical luminosity non-trivially. Although the presence of $q$ and $a_{*}$ does not substantially modify the luminosity from the disk at low frequencies, the behaviour in the high frequency regime is significantly different compared to the Schwarzschild or general relativistic scenario. Thus one can conclude that the presence of both $q$ and $a_{*}$ affect the luminosity appreciably. For a fixed value of rotation parameter $a_{*}$, a positive $q$ enhances the luminosity, while a negative charge parameter diminishes the luminosity at high energies. Similarly, for a fixed $q$, a positive rotation parameter increases the luminosity compared to a non-rotating or retrograde black hole. i.e., prograde spacetimes lead to greater luminosity from the disk. Therefore, luminosity from the disk escalates with positive values of $q$ and $a_{*}$. It is worth emphasizing that a negative charge parameter is a distinctive feature of extra dimensions and such a feature, if supported by observations, will mark a deviation from \gr. We would also like to point out that $q > 0.25$ is not allowed, since the event horizon would disappear with the formation of a naked singularity. On the other hand, the charge parameter $q$ can assume any negative value. Since the parameter $q$ is inherited from the bulk Weyl tensor, it should affect all the black holes in the brane hypersurface in an identical manner such that the value of $q$ should be the same for all the quasars. The rotation parameter $a_{*}$ on the other hand is a characteristic feature of individual quasars and hence should vary from one quasar to another. We need to take an account of this fact when we compare our theoretical luminosities with observations to constrain $q$ and $a$. In the next section, we will analyze the optical spectra of a sample of 80 quasars to see whether these observations provide plausible indications of extra spatial dimensions.

\section{Numerical analysis: Comparison between theoretical and observed luminosity}\label{Accretion_Obs}

In this section, we compare theoretical estimates of optical luminosity of a sample of eighty Palomar Green quasars with the corresponding observed values. Our interest mainly involves optical luminosity because, for supermassive black holes, the luminosity from the accretion disk generally peaks in the optical part of the spectrum. 
 
The theoretical optical luminosity, given by $L_{\rm opt}\equiv \nu L_{\nu}$, is computed at frequency $\nu$ corresponding to wavelength 4861\AA \cite{Davis:2010uq}. Since we are studying a sample of quasars which are not expected to be edge-on systems, one may invoke a conservative bound on the inclination angle, i.e., $\cos i \in \left(0.5,1\right)$. However, following \cite{Davis:2010uq,Wu:2013zqa} we adopt a typical value of $\cos i \sim 0.8$ in our analysis. We have further verified that for non-rotating black holes, given a fixed tidal charge parameter $q$, the error (e.g., reduced $\chi^2$, Nash-Sutcliffe efficiency, index of agreement test etc.) between the observed and theoretical luminosities gets minimized when $\cos i$ lies between $0.77$ and $0.82$ \cite{Banerjee:2017hzw}, thereby justifying our assumption. Further, the inclination angle of some of the quasars in our sample has been estimated using the degree of polarisation of the scattered radiation from the accretion disk \cite{2017Ap&SS.362..231P}, which is also consistent with our choice of the inclination angle. Following which, we compute the theoretical estimates of optical luminosities for a sample of eighty Palomar Green (PG) quasars studied in \cite{Schmidt:1983hr,Davis:2010uq} using the thin disk approximation for the accretion flow \cite{Novikov:1973kta,Page:1974he} in the background spacetime given by \ref{2-2}. The masses of these quasars have been independently constrained using the technique of reverberation mapping \cite{Kaspi:1999pz,Kaspi:2005wx,Boroson:1992cf,1972ApJ...171..467B,1974ATsir.831....1L,1982ApJ...255..419B,Peterson:2004nu}. For thirteen quasars, the masses are also reported by $M-\sigma$ method \cite{Ferrarese:2000se,Gebhardt:2000fk,Dasyra:2006jy,Wolf:2008sm,Tremaine:2002js}. The bolometric luminosities of these quasars are determined using high quality data in the optical \cite{Neugebauer:1987}, UV \cite{Baskin:2004wn}, far-UV \cite{Scott:2004sv}, and soft X-ray \cite{Brandt:1999cm}. The observed values of the optical luminosities and the accretion rates of these eighty PG quasars are reported in \cite{Davis:2010uq}. Since $M$ and $\dot{M}_{0}$ are known for these quasars and $\cos i$ has already been argued to be $\sim ~ 0.8$, we can vary the metric parameters $q$ and $a_{*}$ to get estimates of the theoretical luminosity that agrees best with the known observed values.

Further, there are restrictions on the values of $q$ and $a_{*}$. This is because, in order to have an event horizon, the radius of the same must be real, which, in dimensionless unit, for the brane world black hole is given by,
\begin{align}
x_{\rm EH}=1+\sqrt{1-a_{*}^2-4q} \label{4-1}
\end{align}
Thus suggests that once a value for $q$ is fixed, the spin parameter $a_{*}$ can only assume values between $-\sqrt{1-4q}\le a_{*}\le \sqrt{1-4q}$. Intriguingly, since the brane world black holes can assume negative values for $q$, the absolute value of the spin parameter $a$ can be greater than unity, which is not possible in the context of \gr. 

As already pointed out the two parameters $q$ and $a_{*}$ are however not at the same footing. The tidal charge parameter $q$ affects all the black holes in the brane in an identical manner and hence should be the same for all the quasars, while the rotation parameter $a_{*}$ should vary from one quasar to another. Thus in order to deduce the most favoured choice for the tidal charge $q$ we adopt the following procedure:
\begin{itemize}

\item We start by fixing a value for the tidal charge $q$, which automatically restricts the range of $a_{*}$, viz, $-\sqrt{1-4q}\le a_{*}\le \sqrt{1-4q}$.

\item We select one quasar from the sample (with known $M$ and $\dot{M}_{0}$) and calculate its theoretical estimate of optical luminosity for the chosen $q$ and all the allowed values of $a_{*}$. The value of $a_{*}$ that best agrees with the observed luminosity, is considered to be the spin of that quasar for the chosen $q$.

\item Keeping the $q$ value fixed, we repeat the above procedure for all the eighty quasars which endows a specific value of $a_{*}$ for each of the quasars in the sample.

\item We now repeat the above three operations for different values of $q\le 0.25$.

\item This ensures that for every value of $q$ we assign a spin for all the eighty quasars that minimizes the error between the observed and the theoretical estimates of optical luminosities.

\end{itemize}
This presents the basic numerical procedure we have followed. However, to determine the value of the tidal charge parameter $q$, where the error between theoretical and observed luminosity is minimized, we have computed several error estimators which we will discuss next.   
 
\subsection{Estimating the most favoured tidal charge parameter from observations}

In this section we discuss several error estimators to find out the most favored model of $q$ that minimizes the error between the theoretical and the observed optical luminosities.
\begin{itemize}

\item {\bf Chi-square $\boldsymbol {\chi^{2}}~$}:~
The chi-square ($\chi ^{2}$) of a distribution is given by:
\begin{align}\label{4-2}
\chi ^{2}(q,\left\lbrace a_k\right\rbrace)=\sum _{i}\frac{\left\{\mathcal{O}_{i}-\mathcal{M}_{i}(q,\left\lbrace a_k\right\rbrace) \right\}^{2}}{\sigma _{i}^{2}}.
\end{align}
In \ref{4-2}, $\{\mathcal{O}_{i}\}$ constitutes the set of observed data with corresponding errors $\{\sigma_{i}\}$ while $\{\mathcal{M}_{i}\left(q,\left\lbrace a_k\right\rbrace\right)\}$ represents the theoretical estimates of the observed luminosity for a given model of $\left(q,\left\lbrace a_k\right\rbrace\right)$, $\left\lbrace a_k\right\rbrace$ being the set of best choice of spin parameters corresponding to all the eighty quasars for a given value of $q$ (see discussion in the previous section). For our sample, since the errors $\sigma_{i}$ associated with the observed optical luminosity $L_{obs}$ are not reported, we assign equal weightage to each observation.

It is important to note that we do not use reduced chi-square $\chi ^{2}_{Red}$, where $\chi ^{2}_{Red}=\chi ^{2}/\nu$, ($\nu$ being the degrees of freedom) as an error estimator, since in this scenario there are restrictions on the values of $q$ and $a$ (see previous section). Such systems are called models with prior and the definition of degrees of freedom becomes non-trivial \cite{Andrae:2010gh} in such cases. Since our model falls in such a class we adhere to $\chi ^{2}$ as an error estimator.

The value of $\left(q,\left\lbrace a_k\right\rbrace\right)$ which minimizes the $\chi ^{2}$ is the one that is most favored by observations. In \ref{Fig_2}, the variation of $\chi ^{2}$ with the tidal charge parameter $q$ is shown, which clearly ilustrates that $\chi ^{2}$ attains a minimum for a negative value of $q\sim -0.2$. Since, negative charge parameters are not allowed in \gr, this may signal some new physics at play in the strong gravity regime, higher dimensions being one such possibility. The spin of the quasars that best explain the data are given by the set $\left\lbrace a_k\right\rbrace$ corresponding to $q\sim -0.2$. These spins are reported in \ref{Table1}.

Before discussing more on the spin of the quasars we consider a few more error  estimators, namely the Nash-Sutcliffe Efficiency, the index of agreement and the modified versions of the last two in order to confirm the robustness of our results. 
\begin{figure}[t!]
\centering
\includegraphics[scale=0.75]{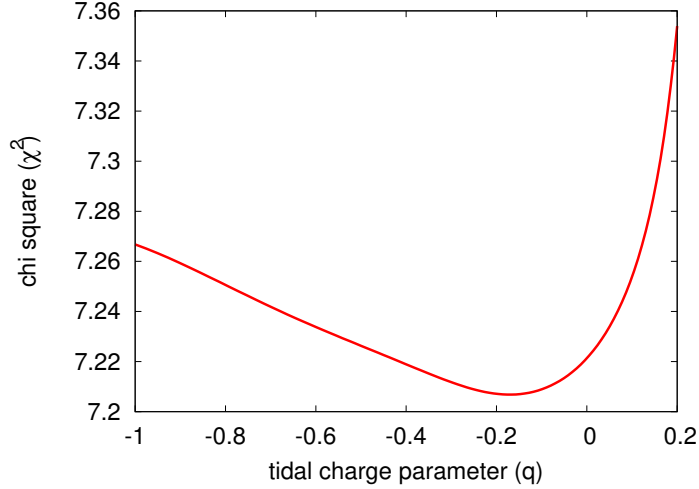}
\caption{Figure 2: The figure illustrates the variation of $\chi^{2}$ with the tidal charge parameter $q$ for a sample of eighty quasars. The minimum of $\chi ^{2}$ corresponds to a negative value of $q\sim -0.2$, which implies that higher dimensional models are favored by observations. For more discussions see text.}
\label{Fig_2}
\end{figure}
\begin{figure}[htp]
\subfloat[Nash Sutcliffe Efficiency \label{Fig_3a}]{\includegraphics[scale=0.65]{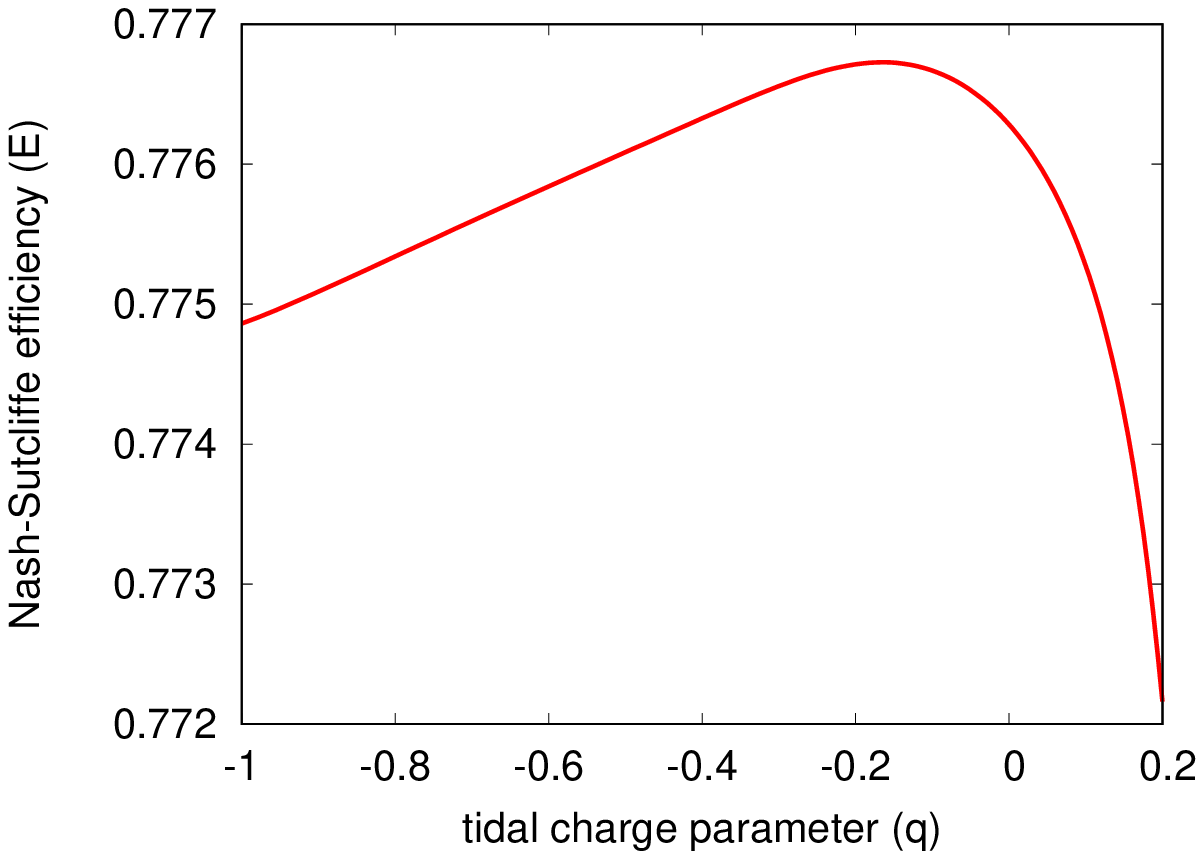}}
\subfloat[Modified form of Nash Sutcliffe Efficiency\label{Fig_3b}]{\includegraphics[scale=0.65]{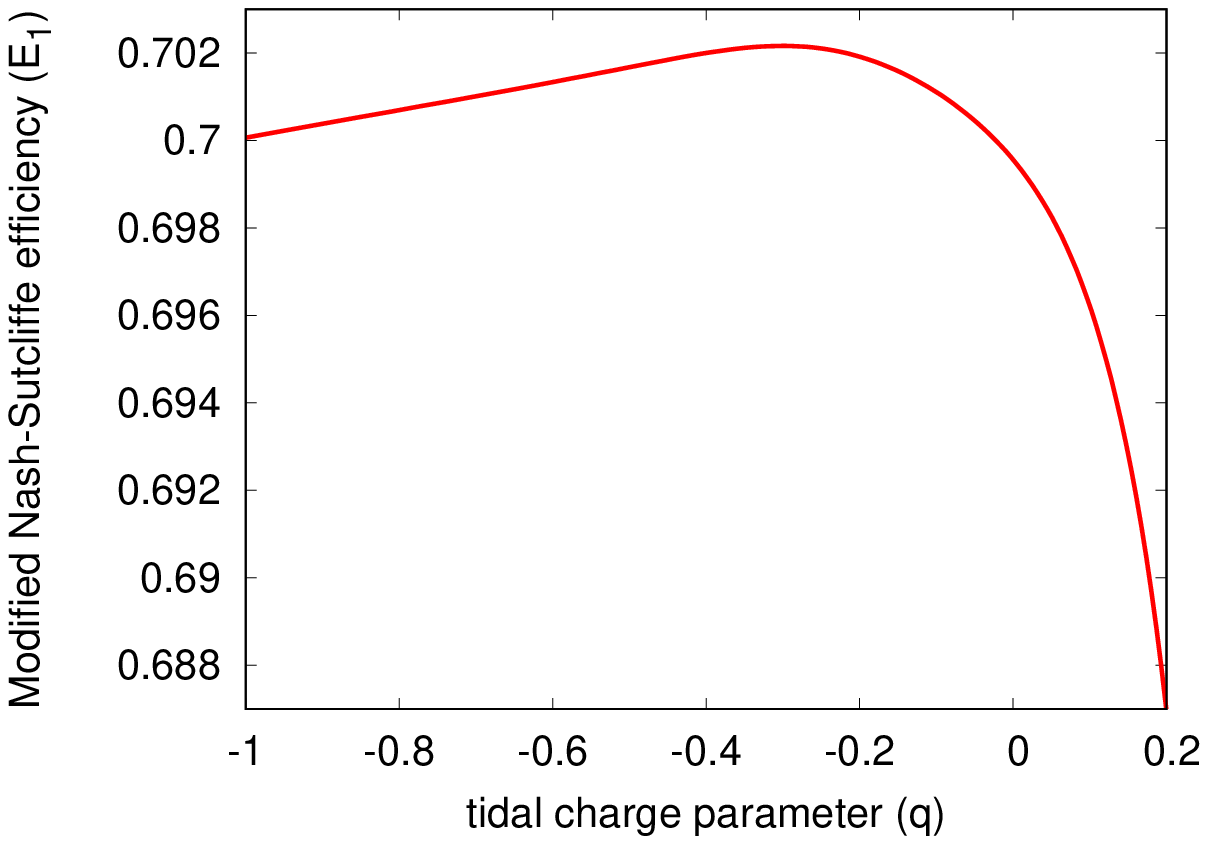}}
\caption{Figure 3: The above figure depicts variation of (a) the Nash-Sutcliffe Efficiency $E$ and (b) the modified form of the Nash-Sutcliffe Efficiency $E_1$ with the tidal charge parameter $q$. Both the error estimators maximize for negative values of $q$.}
\end{figure}

\item \textbf{Nash-Sutcliffe Efficiency and its modified form:} Nash-Sutcliffe Efficiency $E$ \cite{NASH1970282,WRCR:WRCR8013,2005AdG.....5...89K} relates the sum of the squared differences between the observed and the predicted values normalized by the variance of the observed values. Mathematically it is given by,
\begin{align}\label{4-3}
E(q,\left\lbrace a_k\right\rbrace)=1-\frac{\sum_{i}\left\{\mathcal{O}_{i}-\mathcal{M}_{i}(q,\left\lbrace a_k\right\rbrace)\right\}^{2}}{\sum _{i}\left\{\mathcal{O}_{i}-\mathcal{O}_{\rm av}\right\}^{2}}
\end{align}
where $\mathcal{O}_{\rm av}$ denotes average of the observed values of the optical luminosities of the quasars. 

It is important to note that $E$ can assume values from $-\infty ~\rm to ~ 1$. A model with negative $E$ implies that the average of the observed data is a better predictor than the model. Unlike $\chi^{2}$, the most favored model maximises the Nash-Sutcliffe Efficiency. A model which has $E$ very close to $1$ is considered to be an ideal model that predicts the observations with great accuracy. The variation of $E$ with $q$ is depicted in \ref{Fig_3a}. We note that $E$ assumes the maximum value for $q\sim-0.2$, which is very close to the value of $q$ where $\chi ^{2}$ attained a minimum. The maximum value of Nash-Sutcliffe Efficiency corresponds to $E\sim 0.777$ which is indicative of a satisfactory model representing the data \cite{Goyal}. 

A modified version of the Nash-Sutcliffe efficiency $E_1$ is often used to circumvent the oversensitivity of Nash-Sutcliffe efficiency to higher values of the luminosity. This oversensitivity arises for taking square of the error in the numerator of the Nash-Sutcliffe efficiency (see e.g. \ref{4-3}) \cite{WRCR:WRCR8013}. The modified Nash-Sutcliffe efficiency $E_1$ is therefore taken to be,
\begin{align}\label{4-4}
E_{1}(q)&=1-\frac{\sum_{i}|\mathcal{O}_{i}-\mathcal{M}_{i}(q,\left\lbrace a_k\right\rbrace)|}{\sum _{i}|\mathcal{O}_{i}-\mathcal{O}_{\rm av}|}
\end{align}
which increases the sensitivity of this estimator to lower luminosity values as well. Similar to $E$, a model which maximizes $E_1$ is considered to be a better representation of the data. \ref{Fig_3b} illustrates the variation of modified Nash-Sutcliffe efficiency with $q$. $E_1$ maximizes at $q\sim -0.25$ which further corroborates our previous findings. It is important to note that the signature of $q$ is important here and not its exact value since negative tidal charge parameters does not arise in standard general relativistic situations.

\begin{figure}[htp]
\subfloat[Index of agreement \label{Fig_4a}]{\includegraphics[scale=0.65]{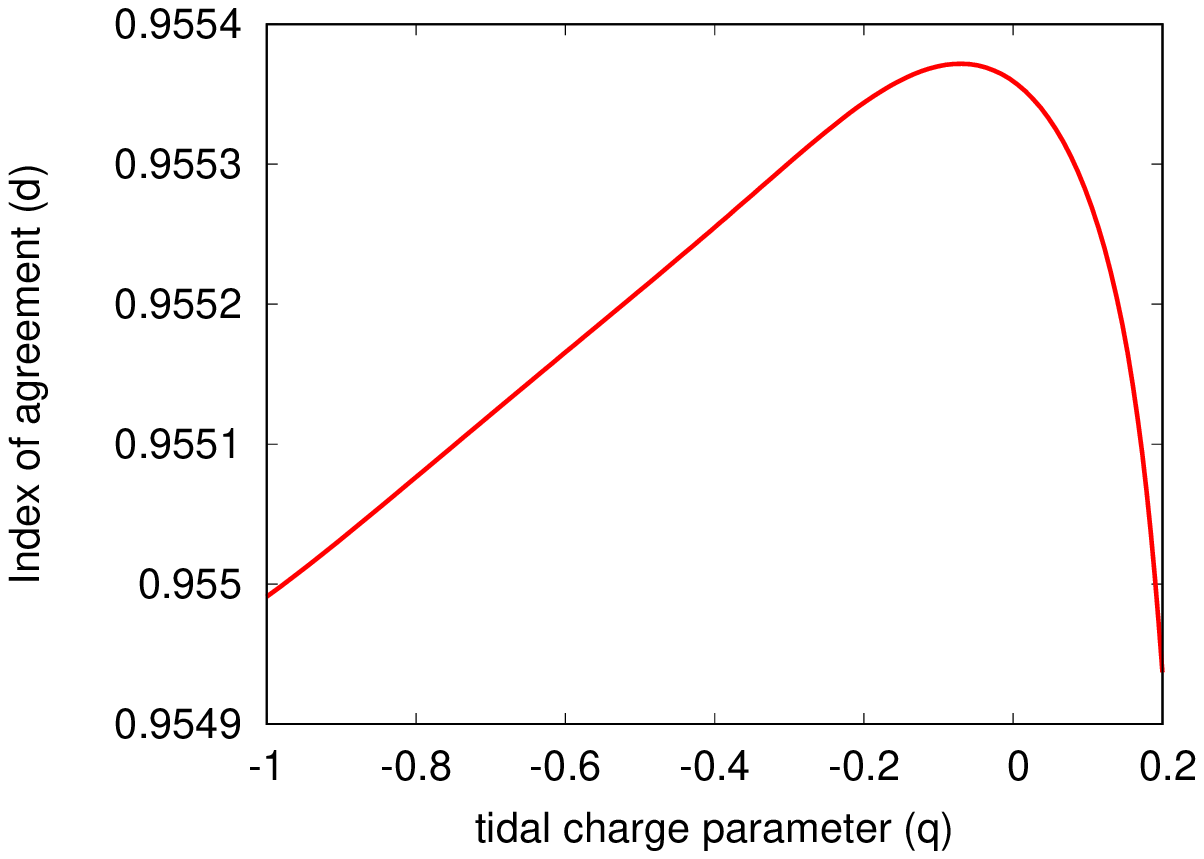}}
\subfloat[Index of agreement\label{Fig_4b}]{\includegraphics[scale=0.65]{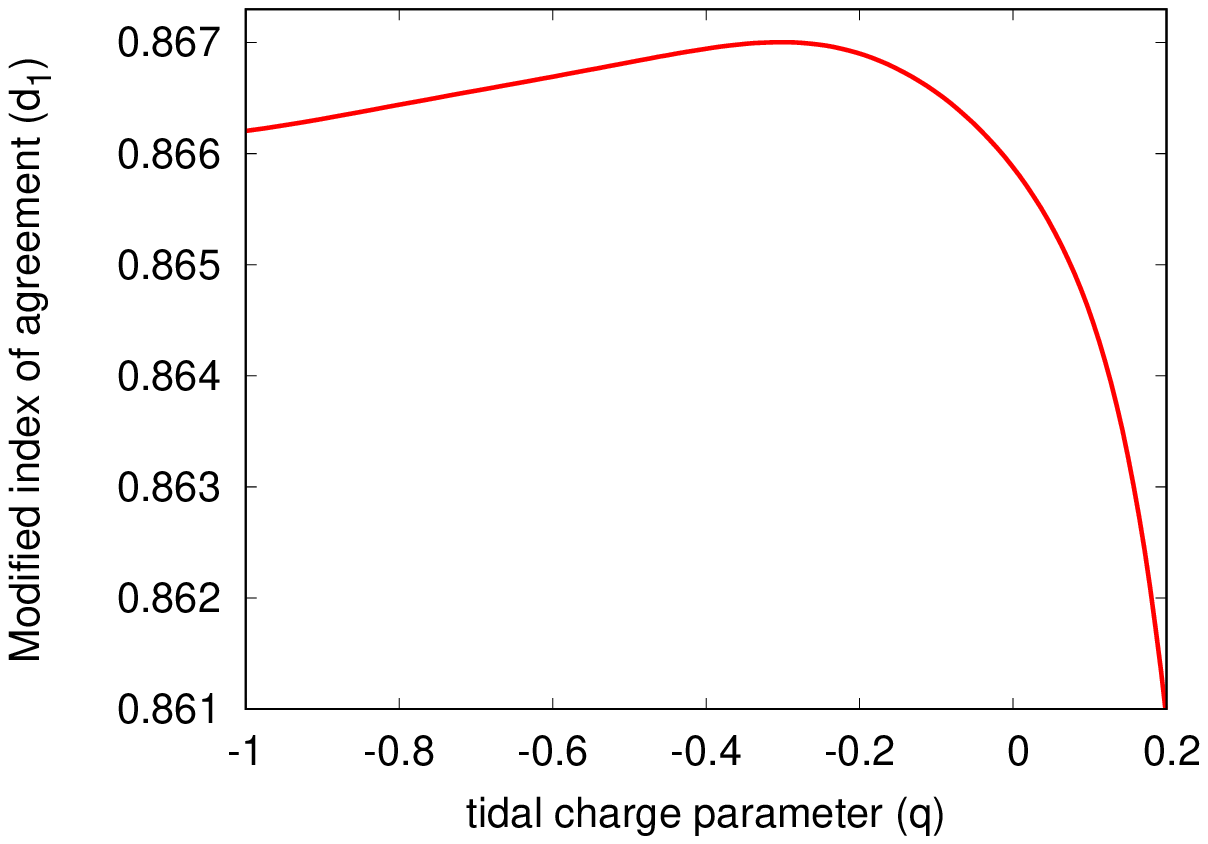}}
\caption{Figure 4: The above figure depicts variation of (a) the Nash-Sutcliffe Efficiency $E$ and (b) the index of agreement $d$ with the tidal charge parameter $q$. Both the error estimators maximize for negative values of $q$.}
\label{Fig_4}
\end{figure}

\item \textbf{Index of agreement and its modified form:}  The index of agreement, denoted by $d$ was proposed by \cite{willmott1984evaluation, doi:10.1080/02723646.1981.10642213,2005AdG.....5...89K} in order to evade the insensitivity associated with the Nash-Sutcliffe efficiency towards the differences between the observed and the predicted luminosities from the corresponding observed mean \cite{WRCR:WRCR8013}. Mathematically, it assumes the form,
\begin{align}\label{4-4}
d(q,\left\lbrace a_k\right\rbrace)=1-\frac{\sum_{i}\left\{\mathcal{O}_{i}-\mathcal{M}_{i}(q,\left\lbrace a_k\right\rbrace)\right\}^{2}}{\sum _{i}\left\{|\mathcal{O}_{i}-\mathcal{O}_{\rm av}|+|\mathcal{M}_{i}(q,\left\lbrace a_k\right\rbrace)-\mathcal{O}_{\rm av}|\right\}^{2}}
\end{align}
where $\mathcal{O}_{\rm av}$ refers to the average value of the observed luminosities.
The quantity in the denominator, known as the potential error, denotes the maximum value by which each pair of observed and predicted luminosities may differ from the average luminosity. 
Similar to Nash-Sutcliffe efficiency, a maximum of $d$ indicates the best model for $q$. \ref{Fig_4a} shows the variation of the index of agreement $d$ with $q$. Here also $d$ maximizes at a negative value of $q$, namely $q\sim -0.1$ which is consistent with the predictions by previous error estimators. 

Similar to Nash-Sutcliffe efficiency, the index of agreement is also oversensitive to higher values of optical luminosity and hence a modified form of this estimator, denoted by $d_1$, is often invoked,
\begin{align}\label{4-5}
d_{1}(q,\left\lbrace a_k\right\rbrace)&=1-\frac{\sum_{i}|\mathcal{O}_{i}-\mathcal{M}_{i}(q,\left\lbrace a_k\right\rbrace)|}{\sum _{i}\left\{|\mathcal{O}_{i}-\mathcal{O}_{\rm av}|+|\mathcal{M}_{i}(q,\left\lbrace a_k\right\rbrace)-\mathcal{O}_{\rm av}| \right\}}
\end{align} 
As before, the model which maximizes $d_1$ is considered to be the more favored model. \ref{Fig_4b} represents the variation of modified index of agreement with $q$. $d_1$ attains a maximum at $q\sim -0.25$. This in turn further strengthens our earlier findings that indeed a \emph{negative} value of $q$ is favored from quasar optical data which signifies the importance of a higher dimensional scenario in the strong gravity regime. 

\end{itemize}
We have already obtained indications of this in an eariler work, \cite{Banerjee:2017hzw} where we considered a spherically symmetric scenario with the spacetime similar to the \RN background with the charge parameter assuming both signs. The same sample of quasars were considered and they were assumed to be non-rotating in that case. Evaluation of the error estimators discussed above revealed that a \emph{negative} charge parameter was favored by observations compared to the Schwarzschild or standard \RN scenario. Since quasars are rotating in general, the current work is aimed to incorporate the spins of quasars by considering the accretion in the corresponding axi-symmetric background.

\subsection{Constraints on the spin of the quasars}
Having established the fact that a negative tidal charge parameter is favored by optical observations of quasars, we now present the estimates of the spins of the quasars from the most favored model of $q$. The procedure for determining the spin has already been discussed in \ref{Accretion_Obs}. Since $q\sim -0.2$ turns out to be the most favored model by observations, the set of spin parameters assigned to the quasars corresponding to $q\sim -0.2$, seems to be the best estimates of spin for the quasars. 

It may seem that the theoretical estimate of optical luminosity $L_{opt}$ depends not only on the radius of the marginally stable circular orbit $r_{ms}$ but also on the outer radius of the accretion disk $r_{out}$ (\ref{3-26}). 
The results presented in the last section were obtained by assuming $r_{out}=500 R_g$ for all the quasars, which is just a typical choice \cite{Walton:2012aw,Bambi:2011jq}. 
In order to verify the sensitivity of the results with the choice of $r_{out}$, we compute all the error estimators discussed in the previous section with $r_{out}=1000 R_g$. Such an analysis reveals that all the error estimators more or less replicate the same trend as observed for $r_{out}=500 R_g$, e.g, $\chi^2_{min}$ is achieved at $q\sim -0.25$ while $E_{max}$ and $d_{max}$ occurs at $q\sim -0.2$ and $q\sim -0.1$ respectively. This indicates that our previous conclusion remains unaltered with variation of $r_{out}$. A negative value of $q$ which arises in a higher dimensional scenario is indeed favored by optical observations of quasars.

The choice of $r_{out}$ however affects the values of the error estimators.
With $r_{out}=1000 R_g$, $\chi^{2}_{min}$ assumes the value, $\sim 2.53$ compared to $r_{out}=500 R_g$, while $E_{max}$ and $d_{max}$ attains values $\sim 0.92$ and $\sim0.98$ respectively, compared to their corresponding $r_{out}=500 R_g$ counterparts (see \ref{Fig_2}, \ref{Fig_3a}, \ref{Fig_3b}, \ref{Fig_4a}, \ref{Fig_4b}). 

It is however intriguing to note that although variation of $r_{out}$ does not substantially alter the values of $q$ corresponding to $\chi^{2}_{min}$, $E_{max}$ or $d_{max}$, it does affect the estimates of spin for some quasars. For these quasars, a change in $r_{out}$ (e.g. from $500 R_g$ to $1000 R_g$) causes $\sim$ 5\%-6\% change in the disk luminosity for a given choice of $q$ and $a$. Therefore, the theoretical luminosity assumes minimum deviation from the observed luminosity for a different value of $a$ if $r_{out}$ is changed from $500 R_g$ to $1000 R_g$. Note that, apart from $q$ and $a$, the theoretical estimate of optical luminosity $L_{opt}$ depends on $M$ and $\dot{M_0}$ which are different for different quasars. For a given $q$ and $a$, the ratio $\dot{M_0}/M^2$ determines how sharply peaked the temperature profile is near $r_{ms}$ (see \ref{3-25} and \ref{3-26}), since the luminosity of a quasar with a sharply peaked temperature profile $T(r)$ near $r_{ms}$ will not have much contribution from the outer disk and hence the choice of $r_{out}$ will not substantially affect the disk luminosity for such quasars. 
This feature actually exhibits a limitation in our method of determining the spin, since the angular momentum of quasars should not be affected by the physical extent of the accretion disk. We therefore report the spin of only those quasars whose luminosity and consequently the spin remains unaltered by variation of $r_{out}$. The best estimates of spin corresponding to $q\sim -0.2$ (the most favored model) and the $q\sim 0$ (for comparison with \gr) has been reported in \ref{Table1}. Note that the spins of quasars reported in the literature till date are based on general relativity and are highly model dependent \cite{Brenneman:2013oba,Reynolds:2013qqa,Reynolds:2019uxi} which we discuss next.

It is apparent from \ref{Table1} that more than half of the quasars in the table are maximally spinning which turns out to be $a_{max}\sim 1.34$ for our higher dimensional model (with $q\sim -0.2$) while $a_{max} \sim 0.998$ \cite{Thorne:1974ve}, if a model based on general relativity is assumed.
This is consistent with the study of Crummy et al. \cite{Crummy:2005nj}, who investigated the spectra of several quasars reported in Table 1 (PG 0003+199, PG 0050+124, PG 0844+349, PG 1115+407, PG 1211+143, PG 1244+026, PG 1402+261, PG 1404+226, PG 1440+356), using the general relativistic disk reflection model \cite{Ross:2005dm} and arrived at a similar conclusion. 
The spins of some of the quasars reported in \ref{Table1} have been independently estimated \cite{Afanasiev:2018dyv} by investigating the polarimetric observations of AGNs. It turns out that our estimate of spins for PG 0003+199, PG 0026+129, PG 0050+124, PG 0844+349, PG 0923+129, PG 0923+201, PG 2130+099 and PG 2308+098 are consistent with their estimates while the results obtained for PG 0921+525, PG 1022+519, PG  1425+267, PG 1545+210, PG 1613+658, PG 1704+608 are not quite similar. 
However, the spin of PG 1704+608 (3C 351), obtained from the correlation between the jet power with the black hole mass and spin \cite{Daly:2013uga} turns out to be similar with our estimates of $a\sim 0.1$, assuming \gr.    
In particular, the spin of PG 0003+199 (also known as Mrk335) has been constrained very precisely to $a \sim 0.89\pm 0.05$ \cite{Keek:2015apa} and $a\sim 0.83^{+0.09}_{-0.13}$ \cite{Walton:2012aw}, by studying its X-ray reflection spectrum and the relativistic broadening of the Fe-$\rm K_{\alpha}$ line, considering general relativity in the strong gravity regime. 
 
Although their method of constraining the spin is different from us, their estimates are well in agreement with our predictions of a maximally spinning black hole in PG 0003+199. 
By investigating the gravitationally red-shifted iron line of PG 1613+658 (Mrk 876) Bottacini et al. \cite{Bottacini:2014lva} inferred that the quasar harbors a rotating central black hole which is in agreement with our results. 
 
The spin of radio-loud quasars, PG 1226+023 (3C 273), PG 1704+608 (3C 351) and PG 1100+772 \cite{Sikora:2006xz,Vasudevan:2007hz}, have been constrained by  Piotrovich et al. \cite{2017Ap&SS.362..231P} to $a<1$, $a<0.998$ and $a\sim 0.88^{+0.02}_{-0.03}$ respectively. According to our model, PG 1226+023 harbors a maximally spinning retrograde black hole while PG 1704+608 and PG 1100+772 are slowly rotating prograde systems. It has been recently proposed \cite{Reynolds:2006uq,Garofalo:2010hk,Garofalo:2009ki,Garofalo:2013ula} that retrograde black holes often harbor strong radio jets by the combined effect of Blandford-Znajeck \cite{Blandford:1977ds} and Blandford-Payne \cite{Blandford:1982di} mechanisms which seems to be consistent with the high radio luminosity observed in these systems \cite{Polletta:2000gi,Vasudevan:2007hz}. Conversely, rapidly spinning prograde systems are generally radio-quiet which seems to be nearly in agreement with our findings \cite{Kellermann:1989tq,Barvainis:2004wr,Sikora:2006xz,Villforth:2009eq}. 

It is important to note that quasars are complicated systems with their spectral energy distribution (SED) containing emissions from various components e.g., the accretion disk, the corona, the jets, the dust torus etc. which are not always easy to observe and model. Disentangling the effects of one from the other is often very challenging that limits accurate determinination of their mass, distance and inclination. This in turn results in a diversity in the spin estimations for the same quasar by different methods \cite{Brenneman:2013oba,Reynolds:2013qqa,Steiner:2012vq,Gou:2009ks,Reynolds:2019uxi}. 
\\
    
\begin{table}[H]
\vskip0.2cm
{\centerline{\large Table 1}}
{\centerline{Spins of supermassive BHs corresponding to $q=-0.2$ and $q=0$ (for comparison with \gr)}}
\caption{}
\label{Table1}
{\centerline{}}
\begin{center}
\begin{tabular}{|c|c|c|c|c|c|c|}

\hline
$\rm Object$ & $\rm log~ m$ & $\rm log ~\dot{m}$ & $\rm log ~L_{obs}$ & $\rm log ~L_{bol}$ & $a_{q=-0.2}$ & $a_{q=0}$\\
\hline 
$\rm 0003+158$ & $\rm 9.16$ & $\rm 0.79$ & $\rm 45.87$ & $\rm 46.92 \pm 0.25$ & $\rm -0.2 $ & $\rm -0.5 $\\ \hline
$\rm 0003+199$ & $\rm 6.88$ & $\rm -0.06$ & $\rm 43.91$ & $\rm 45.13 \pm 0.35$ & $\rm 1.34 $ & $\rm 0.99 $\\ \hline
$\rm 0026+129$ & $\rm 7.74$ & $\rm 0.80$ & $\rm 44.99$ & $\rm 46.15 \pm 0.29$ & $\rm 1.34 $ & $\rm 0.99 $\\ \hline
$\rm 0043+039$ & $\rm 8.98$ & $\rm 0.36$ & $\rm 45.47$ & $\rm 45.98 \pm 0.02$ & $\rm 0.0 $ & $\rm -0.3 $\\ \hline
$\rm 0050+124$ & $\rm 6.99$ & $\rm 0.58$ & $\rm 44.41$ & $\rm 45.12 \pm 0.04$ & $\rm 1.34 $ & $\rm 0.99 $\\ \hline
$\rm 0844+349$ & $\rm 7.50$ & $\rm -0.01$ & $\rm 44.31$ & $\rm 45.40 \pm 0.28$ & $\rm 1.34 $ & $\rm 0.99 $\\ \hline
$\rm 0921+525$ & $\rm 6.87$ & $\rm -0.55$ & $\rm 43.56$ & $\rm 44.47 \pm 0.14$ & $\rm 1.34 $ & $\rm 0.99 $\\ \hline
$\rm 0923+129$ & $\rm 6.82$ & $\rm -0.49$ & $\rm 43.58$ & $\rm 44.53 \pm 0.15$ & $\rm 1.34 $ & $\rm 0.99 $\\ \hline
$\rm 0923+201$ & $\rm 8.84$ & $\rm -0.47$ & $\rm 44.81$ & $\rm 45.68 \pm 0.05$ & $\rm 0.6 $ & $\rm 0.3 $\\ \hline
$\rm 1001+054$ & $\rm 7.47$ & $\rm 0.59$ & $\rm 44.69$ & $\rm 45.36 \pm 0.12$ & $\rm 1.34 $ & $\rm 0.99 $\\ \hline
$\rm 1011-040$ & $\rm 6.89$ & $\rm 0.17$ & $\rm 44.08$ & $\rm 45.02 \pm 0.23$ & $\rm 1.34 $ & $\rm 0.99 $\\ \hline
$\rm 1022+519$ & $\rm 6.63$ & $\rm -0.36$ & $\rm 43.56$ & $\rm 45.10 \pm 0.39$ & $\rm 1.34 $ & $\rm 0.99 $\\ \hline
$\rm 1048-090$ & $\rm 9.01$ & $\rm 0.30$ & $\rm 45.45$ & $\rm 46.57 \pm 0.32$ & $\rm 0.20 $ & $\rm -0.1 $\\ \hline
$\rm 1049-006$ & $\rm 8.98$ & $\rm 0.34$ & $\rm 45.46$ & $\rm 46.29 \pm 0.15$ & $\rm 0.10 $ & $\rm -0.2 $\\ \hline
$\rm 1100+772$ & $\rm 9.13$ & $\rm 0.29$ & $\rm 45.51$ & $\rm 46.61 \pm 0.25$ & $\rm 0.40 $ & $\rm 0.1 $\\ \hline
$\rm 1115+407$ & $\rm 7.38$ & $\rm 0.49$ & $\rm 44.58$ & $\rm 45.59 \pm 0.21$ & $\rm 1.34 $ & $\rm 0.99 $\\ \hline
$\rm 1119+120$ & $\rm 7.04$ & $\rm -0.06$ & $\rm 44.01$ & $\rm 45.18 \pm 0.34$ & $\rm 1.34 $ & $\rm 0.99 $\\ \hline
$\rm 1126-041$ & $\rm 7.31$ & $\rm -0.02$ & $\rm 44.19$ & $\rm 45.16 \pm 0.28$ & $\rm 1.34 $ & $\rm 0.99 $\\ \hline
$\rm 1211+143$ & $\rm 7.64$ & $\rm 0.68$ & $\rm 44.85$ & $\rm 46.41 \pm 0.50$ & $\rm 1.34 $ & $\rm 0.99 $\\ \hline
$\rm 1216+069$ & $\rm 9.06$ & $\rm 0.51$ & $\rm 45.62$ & $\rm 46.61 \pm 0.28$ & $\rm 0.0 $ & $\rm -0.3 $\\ \hline
$\rm 1226+023$ & $\rm 9.01$ & $\rm 1.18$ & $\rm 46.03$ & $\rm 47.09 \pm 0.24$ & $\rm -1.3 $ & $\rm -1.0 $\\ \hline
$\rm 1244+026$ & $\rm 6.15$ & $\rm 0.15$ & $\rm 43.70$ & $\rm 44.74 \pm 0.22$ & $\rm 1.34 $ & $\rm 0.99 $\\ \hline
$\rm 1402+261$ & $\rm 7.64$ & $\rm 0.63$ & $\rm 44.82$ & $\rm 46.07 \pm 0.27$ & $\rm 1.34 $ & $\rm 0.99 $\\ \hline
$\rm 1404+226$ & $\rm 6.52$ & $\rm 0.55$ & $\rm 44.16$ & $\rm 45.21 \pm 0.26$ & $\rm 1.34 $ & $\rm 0.99 $\\ \hline
$\rm 1416-129$ & $\rm 8.74$ & $\rm -0.21$ & $\rm 44.94$ & $\rm 45.82 \pm 0.23$ & $\rm 0.3 $ & $\rm 0.0 $\\ \hline
$\rm 1425+267$ & $\rm 9.53$ & $\rm 0.07$ & $\rm 45.55$ & $\rm 46.35 \pm 0.20$ & $\rm 0.9 $ & $\rm 0.5 $\\ \hline
$\rm 1426+015$ & $\rm 8.67$ & $\rm -0.49$ & $\rm 44.71$ & $\rm 45.84 \pm 0.24$ & $\rm 0.5 $ & $\rm 0.2 $\\ \hline
$\rm 1440+356$ & $\rm 7.09$ & $\rm 0.43$ & $\rm 44.37$ & $\rm 45.62 \pm 0.29$ & $\rm 1.34 $ & $\rm 0.99 $\\ \hline
$\rm 1512+370$ & $\rm 9.20$ & $\rm 0.20$ & $\rm 45.48$ & $\rm 47.11 \pm 0.50$ & $\rm 0.5 $ & $\rm 0.2 $\\ \hline
$\rm 1519+226$ & $\rm 7.52$ & $\rm 0.18$ & $\rm 44.45$ & $\rm 45.98 \pm 0.41$ & $\rm 1.34 $ & $\rm 0.99 $\\ \hline
$\rm 1535+547$ & $\rm 6.78$ & $\rm -0.01$ & $\rm 43.90$ & $\rm 44.34 \pm 0.02$ & $\rm 1.34 $ & $\rm 0.99 $\\ \hline
$\rm 1543+489$ & $\rm 7.78$ & $\rm 1.18$ & $\rm 45.27$ & $\rm 46.43 \pm 0.25$ & $\rm 1.34 $ & $\rm 0.99 $\\ \hline
$\rm 1545+210$ & $\rm 9.10$ & $\rm 0.01$ & $\rm 45.29$ & $\rm 46.14 \pm 0.13$ & $\rm 0.50 $ & $\rm 0.2 $\\ \hline
$\rm 1552+085$ & $\rm 7.17$ & $\rm 0.56$ & $\rm 44.50$ & $\rm 45.04 \pm 0.01$ & $\rm 1.34 $ & $\rm 0.99 $\\ \hline
$\rm 1613+658$ & $\rm 8.89$ & $\rm -0.59$ & $\rm 44.75$ & $\rm 45.89 \pm 0.11$ & $\rm 0.7 $ & $\rm 0.4 $\\ \hline
$\rm 1704+608$ & $\rm 9.29$ & $\rm 0.38$ & $\rm 45.65$ & $\rm 46.67 \pm 0.21$ & $\rm 0.40 $ & $\rm 0.1 $\\ \hline
$\rm 2130+099$ & $\rm 7.49$ & $\rm 0.05$ & $\rm 44.35$ & $\rm 45.52 \pm 0.32$ & $\rm 1.34 $ & $\rm 0.99 $\\ \hline
$\rm 2209+184$ & $\rm 8.22$ & $\rm -0.98$ & $\rm 44.11$ & $\rm 46.02 \pm 0.47$ & $\rm 0.4 $ & $\rm 0.1 $\\ \hline
$\rm 2308+098$ & $\rm 9.43$ & $\rm 0.22$ & $\rm 45.62$ & $\rm 46.61 \pm 0.22$ & $\rm 0.8 $ & $\rm 0.5 $\\ \hline
\end{tabular}

\end{center}
\end{table}

\section{Concluding Remarks}\label{Accretion_Conc}

In this work we aim to investigate the imprints of higher dimensions in the continuum spectra of quasars which are potential sites to probe strong gravity. The black hole solutions in the 3-brane embedded in a higher dimensional bulk inherits a tidal charge parameter from the bulk Weyl tensor which unlike the well-known \KN solution can assume \emph{negative} values. This turns out to be a distinctive signature of extra dimensions which along with the rotation parameter of the black hole modifies the emitted spectrum from the accretion disk significantly. In a previous work \cite{Banerjee:2017hzw} we reported that optical observations of quasars favor a negative tidal charge parameter, assuming the black holes in the quasars to be non-rotating. Since the black holes in general possess some angular momentum, it is instructive to study the interplay of the charge and the spin parameter on the luminosity from the accretion disk. This in turn will allow us to deduce the deviations incurred in our earlier conclusion by the inclusion of the black hole spin.

In order to explore the effects of the tidal charge parameter on the continuum spectrum from the accretion disk we consider a sample of eighty Palomar Green quasars \cite{Schmidt:1983hr,Davis:2010uq} with well determined masses, accretion rates and bolometric luminosities. For each quasar, we compute the theoretical estimate of luminosity from the accretion disk using the thin-disk approximation and compare it with the corresponding observed value. Several error estimators, namely, the $\chi^{2}$, the Nash-Sutcliffe Efficiency, the index of agreement etc. are evaluated for a more quantitative estimate of our results. This not only enables us to infer about the signature of the most favored charge parameter but also allows us to provide an estimate on the black hole spins. We compare the spins of the quasars reported previously with our estimates and find that they are more or less in agreement. Our analysis reveals that incorporating black hole rotation does not alter our previous conclusion \cite{Banerjee:2017hzw} and indeed quasar optical observations favor a \emph{negative} tidal charge parameter which arises in a higher dimensional scenario.  

It is important to note that for more accurate determination of the black hole spins, a more comprehensive analysis of the SED is required taking into account the contributions from the corona and the jet. This requires implementation of a general accretion flow model which relaxes the thin-disk approximations and hence should successfully reproduce the emissions from the disk and the corona. Modelling the jet would be a further upgradation to the theoretical SED. Such a model will also provide stronger constraints on the signature of the tidal charge parameter.

Apart from difficulties in modelling the SED, there are limitations associated with accurate observations of the innermost regions of the accretion disk which is affected by the strong gravity of the central black hole, since the current X-ray telescopes lack the required resolution \cite{Brenneman:2013oba}. In case of supermassive black holes, probing the strong gravity regime becomes even more challenging due to the difficulties in accurately determining their mass, distance and inclinations. These factors therefore often result in diverse constraints on spin for the same black hole. 

Apart from continuum emission emitted from the accretion disk around black holes, one can also resort to other observations to gain a deeper insight into the nature of strong gravity. The broadened and skewed iron K-$\alpha$ line in the reflection spectrum of black holes \cite{Bambi:2016sac,Ni:2016uik}, the quasi-periodic oscillations observed in the power spectrum of some black holes and neutron stars \cite{Maselli:2014fca,Pappas:2012nt}, pulsar timing arrays \cite{Stairs:2003eg,Cornish:2017oic} and imaging the event horizon of black holes with the techniques of Very Large Baseline Interferometry (VLBI) \cite{Psaltis:2008bb,Bozza:2001xd,Virbhadra:1999nm}, all provide independent tools to probe the strong gravity regime and hence enable us to establish/falsify/constrain several alternate gravity models. Finally, one can also probe the dynamical nature of gravitational interactions by studying the gravitational waves emanating from the mergers of binary black holes using future missions like Laser Interferometer Space Antenna \cite{Will:2004xi,Cardoso:2017cfl,Chakravarti:2018vlt,Chakraborty:2017qve}. We are presently working on the observational aspects of different alternate gravity models which we aim to report in future.
\section*{Acknowledgements}

The research of SSG is partially supported by the Science and Engineering Research Board-Extra Mural Research Grant No. (EMR/2017/001372), Government of India. Research of S.C. is funded by the INSPIRE Faculty Fellowship (Reg. No. DST/INSPIRE/04/2018/000893) from the Department of Science and Technology, Government of India.
\bibliography{KN-ED,Gravity_1_full,Gravity_3_partial,Gravity_2_full,Brane,My_References,axion}

\bibliographystyle{./utphys1}
\end{document}